\documentclass[doublecol]{epl2} 
\usepackage{graphicx}
\usepackage{amssymb}

\title{Manifestation of the spin textures in the thermopower of MnSi}

\author{Stevan Arsenijevi\'c\inst{1} \and Cedomir Petrovic\inst{2} \and L\'aszl\'o Forr\'o\inst{1} \and Ana Akrap\inst{3} }
\shortauthor{S. Arsenijevi\'c \etal}

\institute{
\inst{1} Institut de Physique de la mati\`{e}re complexe,
EPFL, CH-1015 Lausanne, Switzerland \\
\inst{2} Condensed Matter Physics and Materials Sciences Department, Brookhaven National Laboratory, Upton New York, 11973, USA \\
\inst{3} University of Geneva, CH-1211 Geneva 4, Switzerland
}

\pacs{75.30.Kz}{Magnetic phase boundaries}
\pacs{72.15.Jf}{Thermoelectric and thermomagnetic effects}
\pacs{75.50.-y}{Studies of specific magnetic materials}

\abstract{
To identify possible spin texture contributions to thermoelectric transport,  we present a detailed temperature and pressure dependence of thermopower $S$ in MnSi, as well as a low-temperature study of $S$ in a magnetic field. We find that $S/T$ reconstructs the $(p,T)$ phase diagram of MnSi encompassing the Fermi liquid, partially ordered, and non-Fermi liquid phases. Our results indicate that the latter two phases have essentially the same nature. In the partially ordered phase, $S(T)$ is strongly enhanced, which may be understood as a spiral-fluctuation-driven phase. A low temperature upturn in $S/T$ pertaining to the partial order phase persists up to the highest pressure, 24 kbar. Contrarily, a small suppression of $S(T)$ is observed in the ordered skyrmion lattice $A$ phase.}

\begin{document}

\maketitle

The lack of inversion symmetry in MnSi, a $B20$ cubic compound, leads to the presence of intricate magnetic states. A helimagnetic order below $T_c=30$ K results as a compromise between ferromagnetic exchange and Dzalozhinski-Moriya interaction \cite{BakJensen,Nakanishi}. Close to $T_c$, a small magnetic field is sufficient to unpin the helimagnetic order and create a novel magnetic phase, in which long-range order of skyrmions (topologically stable knots of spin structure \cite{PfleidererNV}) persists in a narrow range of magnetic field and temperature, within the $A$ phase of MnSi \cite{Muhlbauer, Rossler, PRLNeubauer}.  
The high pressure behavior of MnSi is intriguing in many aspects. At $p_c\approx $ 14.6 kbar, the helicoidal ferromagnetic state is suppressed and  a mysterious partially ordered (PO) magnetic phase emerges \cite{PfleidererNeutrons}. Simultaneously, an nomalous $T^{3/2}$ dependence of the resistivity sets in, marking the non-Fermi liquid (NFL) behavior that spans a remarkably wide pressure range up to at least 40 kbar \cite{PfleidererNFL,DoironNFL,Pedrazzini}.
Embedded within the large NFL phase, the PO phase was observed by neutron diffraction only in the narrow vicinity of $p_c$ \cite{PfleidererNeutrons}. 
Muon spin relaxation measurements showed that the PO phase is characterized by the absence of a static magnetic order, and indicated that the spin correlations in this phase are dynamic at a timescale between 10$^{-10}$ and 10$^{-10}$~s \cite{UemuraNP}.
PO was interpreted as a phase of fluctuating helices or a blue phase \cite{WrightMermin,PfleidererNeutrons,Hamann}. The link between this disordered spin texture phase and the very wide NFL phase remains unclear. 

In this paper we study the thermoelectric transport in the established spin-textured phases of MnSi: the skyrmion lattice $A$ phase and the high pressure PO phase. We identify an NFL contribution to the thermopower $S$, an enhancement which spans a wide region above the critical pressure $p_c$ and below 10 K. In addition, we distinguish an entropic signature of partial order with the thermopower that anomalously increases in an area in the $(p,T)$ diagram localized around $p_c$, coinciding with the range where PO was detected. Finally, we find a small suppression of $S$ in the $A$ phase, which we attribute to the reduced entropy caused by the formation of an ordered skyrmion lattice. 
Despite the different signatures in the ordered skyrmion $A$ phase and PO phase, our thermopower results suggest that both have their origin in spin textures. The high pressure NFL phase bears similarities to the blue phase \cite{Hamann, TewariPRL} and the single spiral domains \cite{KrugerPRL} suggested by recent theoretical work. Thermopower proves to be an excellent tool to probe spin textures to which the electronic system is strongly coupled.

Single crystals of MnSi were synthesized by Zn flux method \cite{synthesis}.
The high quality of the samples is reflected via a high residual resistivity ratio, RRR$= \rho(300 K)/ \rho_0=300$. 
A sample equipped with four silver paint contacts was mounted on a homemade thermopower sample holder which fits into a non-magnetic clamped pressure cell. Small metallic heaters installed at both ends of the sample generated the temperature gradient measured by a Chromel-Constantan differential thermocouple. Such a setup enabled a simultaneous measurement of the resistivity and thermopower. Daphne oil was used as a pressure medium, with a maximum pressure of 24 kbar. The pressure was determined using a calibrated InSb pressure gauge. 
Orientation of the sample in the magnetic field was along the [100] direction, and the transport measured in the orthogonal plane. 
\begin{figure}[htb]
\centering
\includegraphics[width=0.9\linewidth]{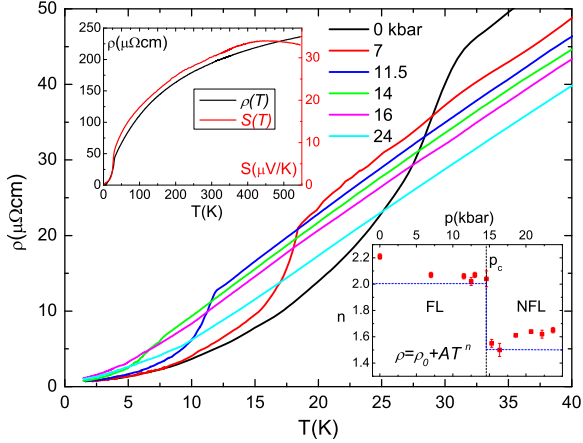}
\caption{(Color online) The temperature dependence of the resistivity for different pressures below and above the critical pressure $p_c=14.6$ kbar. Top inset shows the temperature dependence of thermopower and resistivity up to high temperatures at ambient pressure. Bottom inset shows the power law exponent $n$ extracted from the resistivity curves as a function of pressure. Dashed horizontal lines are guides for the eye, and the vertical line denotes the critical pressure $p_c$.
\label{fig:rho}}\end{figure}

MnSi is a weak itinerant-electron ferromagnet. The compound is a good metal in which the mean free path reaches 5000 \AA \cite{PfleidererNFL}, yet its transport properties have a rather anomalous temperature dependence. The top inset of Fig. \ref{fig:rho} shows the temperature dependence of the resistivity $\rho$ and thermopower $S$ at ambient pressure. The two transport coefficients have a strikingly similar temperature dependence, particularly in the paramagnetic phase. Both $\rho$ and $S$ drop precipitously at ferromagnetic ordering temperature, $T_c$, consistent with a decrease in scattering as an ordered phase is established. 

The resistivity in the paramagnetic phase exhibits nonlinear temperature dependence, contrary to what one would normally encounter in simple metals and attribute to scattering on phonons. Moreover $\rho$ approaches saturation at high temperatures. It has previously been pointed out \cite{MenaPRB} that the resistivity of MnSi can be phenomenologically modeled through a simple fit with two parallel resistors: $1/\rho=1/\rho_\infty+1/(aT)$, up to 300 K. Our data confirms and extends this fit up to 550 K with similar parameters: $\rho_\infty=314\,\mu\Omega$cm and $a=1.7\,\mu\Omega$cmK$^{-1}$. 

$S$, similarly to $\rho$, shows nonlinear temperature dependence in the entire temperature range, and saturates above 400 K. It is hole-like for all temperatures.
The thermopower of a simple metal is given by Mott's formula \cite{Barnard}:
$ S=\pi^2k_B/(3e) T (\partial \ln\sigma/\partial \epsilon)_{\epsilon=\epsilon_F}$, 
where $\sigma$ is conductivity, $\epsilon$ is chemical potential and $\epsilon_F$ the Fermi level.
It is proportional to the variation in $\sigma$ when the chemical potential is infinitesimally shifted \cite{Behnia}. 
The nonlinearity of $S(T)$ in MnSi indicates that $ \partial \ln\sigma / \partial \epsilon$ has significant temperature dependence. Both resistivity and thermopower may reflect scattering on temperature-dependent spin fluctuations.
Indeed, there has been experimental evidence of paramagnetic spin fluctuations taking place up to high temperatures \cite{Ishikawa}; these fluctuations were shown to follow the Moriya-Kawabata theory for the weak itinerant-electron ferromagnet up to $T \sim 10$~$T_c$. Later, this was interpreted as a blue phase of MnSi \cite{Hamann}.

The main panel of Fig. \ref{fig:rho} shows the resistivity at selected pressures. As the pressure increases, $T_c$ is suppressed and can be followed through a kink in $\rho(T)$. The kink becomes progressively sharper until it disappears when $T_c<10$ K. 
Below $T_c$ the resistivity can be described using a power law formula, $\rho(T)=\rho_0+AT^n$. The helicoidal phase at low pressures is characterized by a power law coefficient $n=2$. Above a critical pressure $p_c\approx 14.6$ kbar, the power law exponent suddenly drops to $n=1.5$, as shown in the bottom inset of Fig. \ref{fig:rho}, reproducing the results of Pfleiderer \emph{et al} \cite{PfleidererJLTP}. This is the hallmark of non-Fermi liquid (NFL) behavior in MnSi. The magnetic moments are localized within helical order or the skyrmion lattice \cite{Muhlbauer}, below $T_c$ and $p_c$. In contrast, the low-pressure properties above $T_c$, as well as NFL behavior, reflect the non-localized nature of the magnetic moments.

\begin{figure}[htb]
\centering
\includegraphics[width=0.9\linewidth]{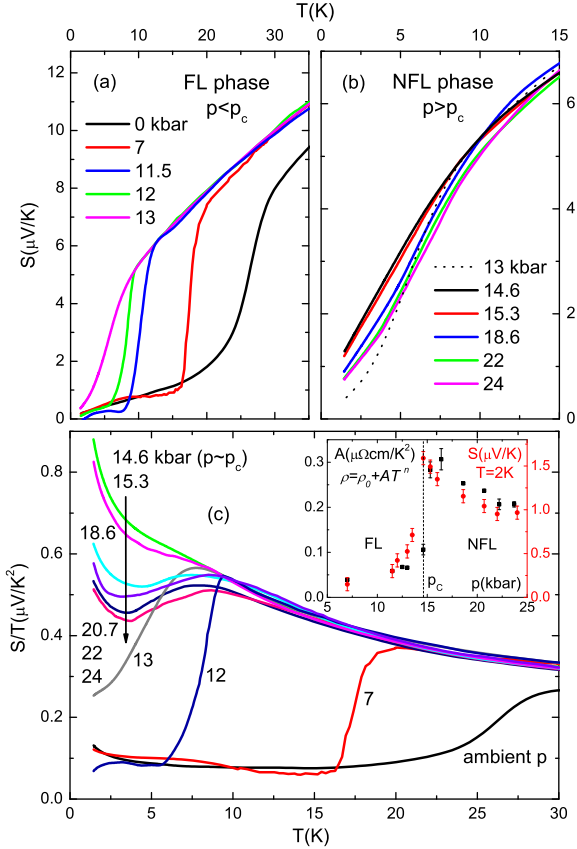}
\caption{(Color online) (a) Temperature dependence of the thermopower $S(T)$ in the FL phase, $p<p_c$. 
The suppression of the ferromagnetic transition can be followed as the drop in $S(T)$ shifts to lower temperatures under pressure. 
(b) $S(T)$ for $p> p_c$, in the NFL phase. Dashed curve for $S$(13 kbar) is shown for comparison. 
(c) $S/T$ as a function of temperature in the FL and NFL phase. 
Inset: pressure dependence of the power law coefficient $A$ (obtained by power law fits of resistivity) and $S(T=2\,\rm K)$.
\label{fig:Sp}}\end{figure}

Signatures of the NFL phase can also be observed in the thermopower, as it is shown in Fig. \ref{fig:Sp},  and previously noted in the work of Cheng \emph{et al} \cite{Cheng}. Because of much lower noise in our thermopower measurements and the high quality of the sample, one can with certainty discern several important additional features.
In the FL phase, shown in Fig. \ref{fig:Sp}a, $S(T)$ behaves similarly to $\rho(T)$. Magnetic ordering is accompanied by a change of the slope and a dramatic decrease in $S(T)$, a consequence of the loss of spin entropy. At low temperatures, $S(T)$ extrapolates to zero in the FL phase. 
The change in slope sharpens as $p$ increases, until $p=12.5$ kbar, where $T_c=10$ K and the drop in $S(T)$ becomes less pronounced than at lower pressures. Even when $T_c$ is entirely suppressed for $p>p_c$ (Fig. \ref{fig:Sp}b), a change in slope of $S(T)$ remains around 10 K. Coincidentally, this temperature is also the upper NFL bound suggested by $\rho(p,T)$.

Above $p_c$ there is a clear enhancement in the low temperature $S(T)$ with respect to the values below $p_c$, shown in Fig. \ref{fig:Sp}b. For $p>p_c$, $S$ no longer extrapolates to zero at zero temperature.
A significant change in $S(T)$ at low temperatures above $p_c$ was previously shown \cite{Cheng}; however, the reported data showed rather different temperature and pressure dependence. 
To illustrate the behavior of $S$ across the quantum phase transition at $p_c$, the inset of Fig. \ref{fig:Sp}c shows the pressure dependence of $S$ at 2 K and the resistivity power law coefficient $A$. $A$ diverges at $p_c$ \cite{PfleidererJLTP} and $S(2\rm\,K)$ shows a very similar behavior. 
As pressure is increased to 24 kbar, the low-temperature contribution to $S(T)$ is gradually suppressed over a wide pressure range. Although the suppression is slow, it still takes place within the phase where the resistivity exponent $n$ decidedly points to NFL behavior. 
This indicates that the increase in $S(T)$ in the vicinity of the critical pressure is not a direct consequence of NFL behavior.

Figure \ref{fig:Sp}c shows a plot of $S/T$ both in the FL and the NFL phase. 
There is a very clear difference between the FL phase, in which at low temperatures $S/T$ is constant, and the NFL region where $S/T$ diverges at low $T$. Similarly, in the established Fermi liquid phases of heavy-fermion systems, $S/T$ is constant, but becomes strongly enhanced in the vicinity of a quantum critical point \cite{HartmannPRL}.   
However, a peculiarity of MnSi, where there is no quantum critical point at $p_c$, is that there is a low-$T$ upturn of $S/T$ that persists even at the highest pressures, far from $p_c$. 
At $p=24$ kbar, $S/T$ has an upturn below 3 K, whereas the NFL behavior sets in at $\sim 10$ K.
With increasing pressure, the slope of the upturn seems to gently decrease and the temperature of its onset is lowered. One can tentatively associate the upturn in $S/T$ with the onset of the PO phase. 

\begin{figure}[htb]
\centering
\includegraphics[width=0.9\linewidth]{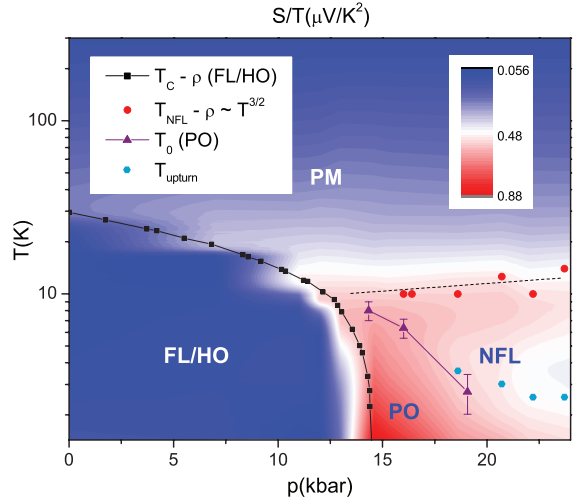}
\caption{(Color online) A contour plot of $S/T$ as a function of temperature and pressure. Superimposed are the pressure dependence of $T_c$  from Doiron-Leyraud et al.\cite{DoironNFL}, $T_{NFL}$ determined from the resistivity (the upper limit of the power law dependence), and $T_0$, a crossover temperature below which partial order was detected using neutron diffraction \cite{PfleidererNeutrons}; in addition we show the temperature $T_{upturn}$ where an upturn starts in $S/T$. Main phases of MnSi are marked: paramagnet (PM), FL/HO (helically ordered Fermi liquid), PO (partial order) and NFL (non Fermi liquid).
\label{fig:Scont}}\end{figure}

To distinguish between the low-$T$ enhancement in $S$, and the NFL response for both the resistivity and thermopower, we plot $S/T$ as a function of temperature and pressure in Fig.~\ref{fig:Scont}. Superposed on top of the contour plot are the pressure dependences of the temperatures $T_c$, $T_{NFL}$ (the upper limit of the power law validity in the resistivity), $T_0$ where partial order was detected, and $T_{upturn}$, the temperature of the onset of an upturn in $S/T$.  
Surprisingly, $S/T$ accurately maps the entire $(p,T)$ phase diagram of MnSi. The dark blue region in the left bottom part of Fig. \ref{fig:Scont} corresponds well to the FL phase: $p<p_c$, $T<T_c$, and $S/T\lesssim 0.48\mu$V/K$^2$. A limit between FL/helically ordered phase and NFL/PO  state can be clearly discerned for $S/T\approx 0.48\mu$V/K$^2$ (white area). This border corresponds to the change in slope in $S$ at around 10~K which can be seen in Fig.~\ref{fig:Sp}b.
The rectangular (red) region above $p_c$ and below 10 K, delimited by $S/T = 0.48\mu$V/K$^2$,  coincides with the NFL border given by the upper limit of the validity of the power law behavior in the resistivity data (Fig.~\ref{fig:rho}).
Finally, nested within the NFL region is a small triangular area just above $p_c$ where $S/T$ reaches its maximum value, $0.88\mu$V/K$^2$. In this part of the phase diagram, the lines of constant $S/T$ are parallel to the edge of the region where neutron scattering experiments observed partial order \cite{PfleidererNeutrons}. The area below $S/T=0.58\mu$V/K$^2$ value the temperature is close to PO crossover temperature $T_0$ identified through neutron diffraction. The enhanced $S/T$ can therefore be taken as the entropic signature of the PO phase. The temperature $T_{upturn}$ where an upturn in $S/T$ takes place above 18~kbar is plotted; this suggests that the PO state continues up to higher pressures.
The above phase diagram suggests that while in $S/T$ we find a fingerprint of the PO phase, there is no phase transition or crossover between PO and NFL. Rather, the two are fundamentally the same phase.

Understanding the PO phase has proven to be a difficult task. PO has been detected only by neutron diffraction experiments \cite{PfleidererNeutrons,PfleidererSANS}, and has remained occult to many other techniques. 
The small angle neutron scattering experiments \cite{PfleidererSANS} indicate that at pressures above $p_c$ there is a mechanism that actively drives the system away from the helical order into a new state. The magnetic field at which helical order is unpinned does not change under pressure up to 21~kbar, suggesting that behavior of MnSi above $p_c$ is not dominated by quantum criticality \cite{PfleidererSANS}.
Several plausible theoretical scenarios for the origin of the PO were put forward. Among them are blue quantum fog \cite{TewariPRL}, blue phase \cite{Hamann}, quantum order by disorder \cite{KrugerPRL}, and amorphous skyrmion phase \cite{Rossler}. 
The main issues to resolve regarding the PO phase are why the ordering vector abruptly shifts from the [111] direction in the helically ordered phase to [110] in the PO phase, and why the anomalous $T^{3/2}$ resistivity behavior persists at pressures far beyond the detected PO. The order by disorder model explains that the change in the ordering direction under pressure may be caused by the leading anisotropy in the spin-orbit coupling \cite{KrugerPRL}. But the question remains as to the large extent of $T^{3/2}$ resistivity and its relation to PO.
The data presented above suggest that the PO phase persists up to higher pressures, similar to the NFL state of MnSi.
Although NFL and PO in MnSi seem to be coupled, it is still an open question how exactly these two phases are related and what is the origin of the observed unusual behavior of the thermopower.
 
\begin{figure}[htb]
\centering
\includegraphics[width=0.9\linewidth]{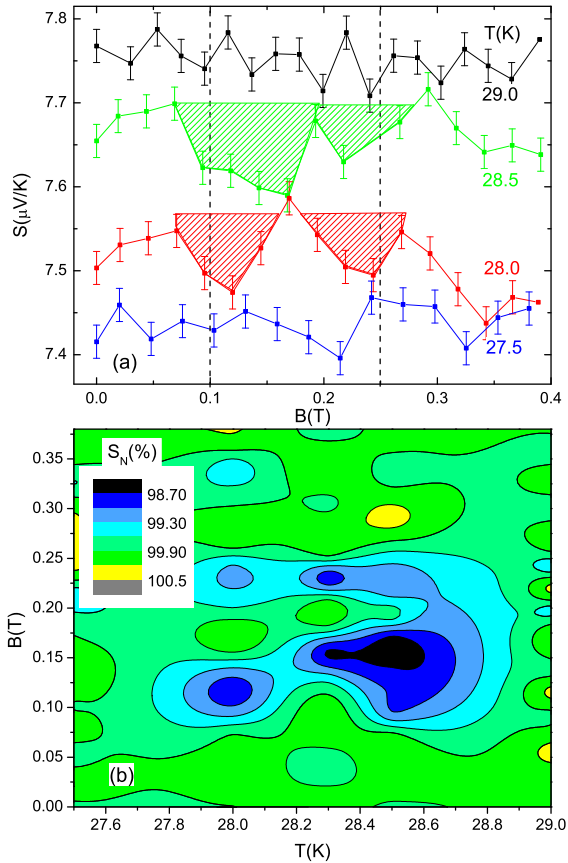}
\caption{(Color online) Top panel: $S(B)$ is shown at four different temperatures at ambient pressure. The thermopower scale shown is that of the data for 29 K, the other traces are offset for clarity. Bottom panel: $S(B,T)$ map in the $A$ phase and its surroundings. \label{fig:SB}}\end{figure}

To better understand what thermopower implies for the possible spin textures in the PO phase, we investigate $S$ in the well-established skyrmion lattice $A$ phase, imaged by neutron diffraction and detected by other probes \cite{Muhlbauer, PRLNeubauer,PRLBauer}. 

The $A$ phase is contained in a small pocket delimited by 0.1 T $<H<$ 0.3 T and 27 K $<T<$ 29 K \cite{MagnetoresistanceJPSJ}. It is surrounded by helicoidal and conical phases, and stabilized through a delicate balance of interactions. 
When a conduction electron traverses a skyrmion, its spin follows the local magnetization and hence continuously changes direction. In this way the electron acquires a Berry phase and experiences an effective Lorentz force, which in turn increases the Hall effect \cite{PRLNeubauer, JonietzScience}. The scattering of conduction electrons may also be affected by an underlying skyrmion lattice. Indeed, small signatures of the $A$ phase were observed in the transport properties of MnSi. Magnetoresistivity is increased by 1\% \cite{MagnetoresistanceJPSJ}, and  the Hall effect enhanced by $\Delta\rho_{xy}\approx 4$ n$\Omega$cm  \cite{PRLNeubauer}. 

Similar effects could also take place when conduction electrons diffuse in the presence of an external thermal gradient. 
We do in fact observe a small decrease in thermopower in the $A$ phase. The top panel of Fig. \ref{fig:SB}  shows the magnetic-field dependence of the thermopower, $S(B)$, for selected four temperatures. $S(B)$ was measured at several different temperatures within and near the $A$ phase. The magnetic field is applied along the $[100]$ direction and perpendicular to the thermal and electrical current flow, so that the transport is measured in the plane of the skyrmion lattice.
While at 27.5 K and 29 K $S(B)$ is rather flat, inside the $A$ phase, at 28 K and 28.5 K, there are two distinct suppressions of $S(B)$ roughly in the interval  0.1 T $<B<$ 0.3 T. A larger set of data ($S(B)$ taken at several different temperatures) gives a map of normalized thermopower $S_N$, shown in a contour plot in the bottom panel of Fig. \ref{fig:SB}.  The values  of $S_N$ are obtained using the expression $S_N=[S_{bf}(B,T)-S(B,T)]/S_{bf}(B,T)$. Here, $S_{bf}$ is baseline thermopower, a smooth background obtained by fitting $S(B,T)$ to a polynomial function for the values of $B$ outside the $A$ phase. The suppression of $S$ seems to occur at the boundaries of the $A$ phase, very similar to the double structure of the $A$ phase seen in the magnetoresistivity measurements \cite{MagnetoresistanceJPSJ}.
The observed suppression of $S_N$ within the $A$ phase is a small effect, less than 1.5\% at its maximum occurring for 28.5 K and 0.15 T ($\Delta S(B,T)\lesssim 0.2\mu$V/K). 

The small amplitude of $\Delta S$ is not surprising when compared to the similarly small changes in magnetoresistivity and the topological Hall effect \cite{MagnetoresistanceJPSJ, PRLNeubauer}; however, its sign is opposite. Ordering a degree of freedom generally leads to a decrease in thermopower since the entropy per charge carrier is lowered. In MnSi, the long range order of the skyrmion lattice leads to a decrease in $S$. If there is also a positive contribution to $S$ due to an enhanced scattering rate of electrons acquiring a Berry phase, it is presumably tiny and compensated by the ordering effect. The net effect is a minute decrease in $S$ that reflects a lowering of entropy.

In contrast to a small decrease in $S$ within the skyrmion $A$-phase, the entropic response of the PO phase is large and positive. The latter may be understood in the following scenario. 
When $p=p_c$, the spin spirals of the helical phase (pinned by the crystal field) reorient and start fluctuating in direction \cite{KrugerPRL}. This fluctuation causes NFL behavior in $\rho$ and produces a fluctuating magnetic structure, possibly an amorphous skyrmion phase \cite{Rossler}. Close to $p_c$, the spiral fluctuations are intense but slow enough that the structure can be detected by neutron scattering, and a large response is seen in $S/T$. When pressure increases, the spin texture fluctuates more freely and hence faster, and the temperature where the fluctuations freeze in decreases, as seen by neutron scattering \cite{PfleidererNeutrons}. A signature of the PO phase is visible through the upturn in $S/T$ and it persists up to the highest pressure measured, $p=24$ kbar. Our results show that PO is an integral part of the NFL phase. This is in accord with the blue phase prediction \cite{Hamann} that PO and NFL are essentially of the same nature; however, the magnetic-correlation length decreases upon increasing pressure in the NFL phase so that the order cannot be detected by NMR or neutron scattering. 

In conclusion, we have identified signatures of both the partially ordered phase and the skyrmion lattice $A$ phase in the thermopower of MnSi. In the PO phase, $S$ shows a strong enhancement that is gradually suppressed as the pressure is increased. $S/T$ accurately maps the $(p,T)$ phase diagram of MnSi, including the PO phase embedded within the NFL phase, in agreement with neutron diffraction \cite{PfleidererNeutrons} and theoretical work \cite{TewariPRL,Hamann,KrugerPRL}. In the $A$ phase a small drop in $S$ signifies a decrease in entropy per carrier, as skyrmions undergo long range order. Thermopower turns out to be an excellent probe of spin textures in MnSi. 

\acknowledgments
We would like to thank A. Rosch, F. Kr\"uger, D. van der Marel and K. Behnia for useful discussions, and N. Miller for helpful comments.  Research was supported by the Swiss NSF and its NCCR MaNEP. Part of this work was carried out at BNL, operated for the U.S. Department of Energy by Brookhaven Science Associates DE-Ac02-98CH10886 (C.P.). A.A. acknowledges funding from ``Boursi\`eres d'Excellence'' of the University of Geneva.


\begin{thebibliography}{1}

\bibitem{BakJensen} P. Bak and M.H. Jensen, J. Phys. C \textbf{13}, L881 (1980).
\bibitem{Nakanishi} O. Nakanishi \emph{et al.}, Solid State Commun. \textbf{35}, 995 (1980).
\bibitem{PfleidererNV} C. Pfleiderer and A. Rosch, Nature \textbf{465}, 880 (2010).
\bibitem{Rossler} U. K. R{\"o}ssler \emph{et al.}, Nature \textbf{442}, 797 (2006).
\bibitem{Muhlbauer} S. M{\"u}hlbauer \emph{et al.}, Science \textbf{323}, 915 (2009).
\bibitem{PRLNeubauer}A. Neubauer \emph{et al.}, Phys. Rev. Lett. \textbf{102}, 186602 (2009).
\bibitem{PfleidererNeutrons} C. Pfleiderer \emph{et al.}, Nature \textbf{427}, 227 (2004). 
\bibitem{PfleidererNFL} C. Pfleiderer, S.R. Julian, and G.G. Lonzarich, Nature \textbf{414}, 427 (2001).
\bibitem{DoironNFL} N. Doiron-Leyraud \emph{et al.}, Nature \textbf{425}, 595 (2003).
\bibitem{Pedrazzini} P. Pedrazzini \emph{et al.}, Physica B \textbf{378Ð380}, 165 (2006).
\bibitem{UemuraNP} Y.J. Uemura \emph{et al.}, Nat. Phys. \textbf{3}, 29 (2007).
\bibitem{WrightMermin} D.C. Wright and N.D. Mermin, Rev. Mod. Phys. \textbf{61}, 385 (1989).

\bibitem{Hamann} Hamann \emph{et al.}, Phys. Rev. Lett. \textbf{107}, 037207 (2011).
\bibitem{TewariPRL} S. Tewari \emph{et al.}, Phys. Rev. Lett. \textbf{96}, 047207 (2006).
\bibitem{KrugerPRL} F. Kr\"uger, U. Karahasanovic, and A.G. Green, Phys. Rev. Lett \textbf{108}, 067003 (2012).
\bibitem{synthesis} R. Hu \emph{et al.}, J. Cryst. Growth \textbf{304}, 114 (2007).
\bibitem{MenaPRB} F. Mena \emph{et al.}, Phys. Rev. B \textbf{67}, 241101 (2003).
\bibitem{Barnard} R. D. Barnard, \emph{Thermoelectricity in Metals and Alloys}, Taylor and Francis Ltd., 
(London, 1972). 
\bibitem{Behnia} K. Behnia, J. Phys.: Condens. Matter \textbf{21}, 113101 (2009).
\bibitem{Ishikawa} Y. Ishikawa \emph{et al.}, Phys. Rev. B \textbf{31}, 5884 (1985).
\bibitem{PfleidererJLTP} C. Pfleiderer, J. Low Temp. Phys. \textbf{147}, 231 (2007).
\bibitem{Cheng} J.-G. Cheng \emph{et al.}, Phys. Rev. B \textbf{82}, 214402 (2010). 
\bibitem{HartmannPRL} See for example S. Hartmann \emph{et al.}, Phys. Rev. Lett. \textbf{104}, 096401 (2010).
\bibitem{PfleidererSANS} C. Pfleiderer \emph{et al.}, Phys. Rev. Lett \textbf{99}, 156406 (2007).

\bibitem{PRLBauer} A. Bauer \emph{et al.}, Phys. Rev. Lett \textbf{110}, 177207 (2013).
\bibitem{MagnetoresistanceJPSJ} K. Kadowaki, K. Okuda, and M Date, J. Phys. Soc. Jpn. \textbf{51}, 2433 (1982). 
\bibitem{JonietzScience} F. Jonietz \emph{et al.}, Science \textbf{330}, 1648 (2010).

\end{thebibliography}
\end{document}